\documentclass[times,twocolumn,final]{elsarticle}

\usepackage{medima}
\usepackage{framed,multirow}
\usepackage{times}
\usepackage{epsfig}
\usepackage{graphicx}
\usepackage{amsmath}
\usepackage{bm}
\usepackage{algorithm}
\usepackage{algorithmic}

\usepackage{amssymb}
\usepackage{latexsym}

\usepackage{url}
\usepackage{xcolor}
\usepackage{multirow}
\usepackage{hyperref}

\definecolor{newcolor}{rgb}{.8,.349,.1}

\journal{Medical Image Analysis}

\begin{document}

\verso{Yuan Xue \textit{et~al.}}

\begin{frontmatter}

\title{
Selective Synthetic Augmentation with HistoGAN for Improved Histopathology Image Classification}

\author[1]{Yuan \snm{Xue}\fnref{fn1}}
\author[1]{Jiarong \snm{Ye}\fnref{fn1}}
\fntext[fn1]{These authors contributed equally to this work.}
\author[1]{Qianying \snm{Zhou}}
\author[2]{L. Rodney \snm{Long}}
\author[2]{Sameer \snm{Antani}}
\author[2]{Zhiyun \snm{Xue}}
\author[2]{Carl \snm{Cornwell}}
\author[3]{Richard \snm{Zaino}}
\author[3]{Keith C. \snm{Cheng}}
\author[1]{Xiaolei \snm{Huang}\corref{cor1}}
\cortext[cor1]{Corresponding author: 
  Xiaolei Huang (sharon.x.huang@psu.edu)}
  
\address[1]{College of Information Sciences and Technology, The Pennsylvania State University, University Park, PA 16802, USA}
\address[2]{Lister Hill National Center for Biomedical Communications, National Library of Medicine, Bethesda, MD 20892, USA}
\address[3]{Department of Pathology, Penn State Health Milton S. Hershey Medical Center and Penn State College Of Medicine, Hershey, PA 17033, USA}


\begin{abstract}
Histopathological analysis is the present gold standard for precancerous lesion diagnosis. The goal of automated histopathological classification from digital images requires supervised training, which requires a large number of expert annotations that can be expensive and time-consuming. Meanwhile, accurate classification of image patches cropped from whole-slide images are essential for standard sliding window based histopathology slide classification methods. To mitigate these issues, we propose a carefully designed conditional GAN model, namely \textit{HistoGAN}, for synthesizing realistic histopathology image patches conditioned on class labels. We also investigate a novel synthetic augmentation framework that \textit{selectively} adds new synthetic image patches generated by our proposed HistoGAN, rather than expanding directly the training set with synthetic images.
By selecting synthetic images based on the confidence of their assigned labels and their feature similarity to real labeled images, our framework provides quality assurance to synthetic augmentation.
Our models are evaluated on two datasets: a cervical histopathology image dataset with limited annotations, and another dataset of lymph node histopathology images with metastatic cancer.
Here, we show that leveraging HistoGAN generated images with selective augmentation results in significant and consistent improvements of classification performance ($\textbf{6.7}\%$ and $\textbf{2.8}\%$ higher accuracy, respectively) for cervical histopathology and metastatic cancer datasets.
\end{abstract}


\end{frontmatter}


\section{Introduction}
\label{Intro}
Image analysis of digitized histopathological slides can contribute significantly to cancer diagnosis~\cite{irshad2013methods}.
For instance, the diagnosis of cervical cancer and its precancerous stages can be accomplished through assessment of histopathology slides of cervical tissue by pathologists. 
An important outcome of the assessment is the 
cervical intraepithelial neoplasia (CIN) grade, an essential indicator for abnormality
assessment identified by the abnormal growth of cells on the surface of the cervix. Over the past decade, computer-assisted diagnosis (CAD) algorithms have been developed for histopathology images to complement the opinion of the pathologist for accurate disease detection, diagnosis, and prognosis prediction~\cite{gurcan2009histopathological}. Considering the shortage of pathologists, automatic histopathology image classification systems have great potential 
in underdeveloped regions for its low cost and accessibility. Moreover, such a system can help pathologists with diagnosis and potentially mitigate the inter- and intra- pathologist variation.

The supervised training of image recognition systems often requires huge amounts of expert annotated data to reach a high level of accuracy. However, for many practical applications using histopathology images, only small datasets of labeled data are available due to annotation cost and privacy concerns, and the labels are often imbalanced between grades and subtypes.
While traditional data augmentation can increase the amount of training data to some degree, commonly employed random transformations or distortions (such as cropping and flipping) lack flexibility and cannot fill the entire data distribution with missing data samples.

Motivated by the aforementioned difficulties in creating sufficiently large training sets for histopathology image recognition systems, we focus on the problem of expanding training sets with high-quality synthetic examples.
Recently, several works in medical image analysis have leveraged unsupervised learning methods, more specifically, Generative Adversarial Networks (GANs)~\cite{goodfellow2014generative}, to mitigate the effects of small training sets on network training
~\cite{liu2019wasserstein,frid2018gan}. These works show that carefully designed GANs can generate visually appealing synthetic images, but two major issues remain insufficiently investigated for generalized and robust synthetic augmentation: 1)
how to mitigate label ambiguity of generated images; and 2) how to ensure the feature quality of synthetic images used for data augmentation.
In other words, blindly incorporating synthetic samples into the original training set, even if they are visually realistic, is not guaranteed to improve the classification model performance. Synthetic images without quality assurance can potentially adversely alter the data distribution and downgrade model performance. We provide a detailed analysis in Section~\ref{sec:results}.

In this paper, we aim at solving these two issues by designing a novel conditional GAN (cGAN)~\cite{mirza2014conditional} framework, termed as HistoGAN, for high-fidelity histopathology image synthesis, then \textit{selectively} adding synthetic samples generated by HistoGAN to the original training set. Our proposed HistoGAN model consists of multiple progressive generation and refinement modules which gradually generate images with better quality. To encourage the diversity of synthetic images, we incorporate the minibatch discrimination~\cite{salimans2016improved} to reduce the closeness between
examples inside a minibatch. Self attention~\cite{vaswani2017attention} is employed to capture relationships between pixels inside an image. Such relationships contain crucial information about histopathology images, including the density distribution of nuclei and color changes in different locations. Class conditional batch normalization~\cite{de2017modulating} and spectral normalization~\cite{miyato2018spectral} are also utilized to stabilize the adversarial training process and improve the quality of synthetic images. Further, during HistoGAN training, we calculate a smoothed version of Fréchet Inception Distance (FID)~\cite{heusel2017gans} score after each epoch of training so that the trained models can be compared and the model with weights that give rise to the best FID score can be selected. Our proposed HistoGAN consistently generates realistic histopathology image patches on two different datasets, which shows the robustness and generality of the model. 

Our proposed selective synthetic augmentation framework consists of two steps. First, we select generated images that can be classified into some class with certainty, by calculating the expectation of predictive entropy of each sample and keeping those samples with relatively low entropy (\textit{i.e.}, high label confidence). Second, we compare the features of real images and synthetic images where the ground truth label of the real images matches the conditional label used to generate the synthetic images, and only select those synthetic images that are sufficiently close to the real-image centroid in feature space. The features of the images are extracted by a feature extractor pre-trained with Monte Carlo dropout (MC-dropout)~\cite{gal2016dropout}. This second step of selection is to ensure that a selected synthetic image indeed belongs to the class that corresponds to the conditional label used to generate it.
The total number of selected samples is determined according to the augmentation ratio $r$ (\textit{i.e.}, the proportion of the number of augmented samples to the number of original training samples). Experimental results show that our proposed HistoGAN model along with selective synthetic augmentation significantly outperforms the baseline ResNet34~\cite{he2016deep} model with traditional augmentation, and also outperforms the synthetic augmentation methods without selection.

To validate the effectiveness and generality of our proposed selective synthetic augmentation framework, we conduct extensive experiments on two histopathology datasets. We first study the 4-class (Normal, CIN 1-3) cervical histopathology image classification problem and evaluate our models on a heterogeneous epithelium image dataset~\cite{xue2019synthetic} with limited and highly unbalanced numbers of patch-level annotations per class label. The second dataset we use is a small subset of the PCam dataset~\cite{veeling2018rotation}, consisting of lymph node histopathology images.
We compare our proposed selective synthetic augmentation method with baseline methods including baseline classification models, models trained with traditional augmentation, and models trained with synthetic augmentation but without quality-assuring selection. Experimental results show that our model achieves significant improvements with $\textbf{6.7}\%$ and $\textbf{2.8}\%$ higher accuracy than baseline classification models on cervical and lymph node datasets, respectively.

The main contributions of this work are as follows:
\begin{itemize}
\setlength{\itemsep}{-0.2em}
    \item We design a novel conditional GAN model architecture for synthesizing realistic histopathology image patches. 
    A smoothed version of FID score is used as a metric to select the best cGAN model during training. With only a limited amount of training data, our GAN model can generate synthetic images with high fidelity and diversity. 
    \item We propose a selective synthetic augmentation method that actively selects synthetic samples with high confidence of matching to their conditional label and are close to real images in feature space. By only adding selected synthetic samples instead of arbitrary synthetic samples to augment the limited training set, our proposed method can significantly outperform other baseline augmentation methods in improving classification performance. 
    The proposed selective synthetic augmentation is general and can also be used in conjunction with other augmentation methods.
    \item We conduct extensive experiments on both a cervical histopathology dataset and a lymph node histopathology dataset. Compared with baseline models, including our previous state-of-the-art synthetic augmentation model~\cite{xue2019synthetic}, our proposed method improves the augmented classification performance.
\end{itemize}

\section{Related Work}
\subsection{Histopathology Image Classification}
Machine learning, especially deep learning methods have achieved promising results on general histopathology image classification. While whole slide images (WSI) are often with unusually high resolutions, commonly used methods~\cite{hou2016patch, xu2017large, tomita2019attention} alleviate this issue by applying patch-level image classification on cropped image patches or sliding windows rather than the original WSI. Individual classification results on cropped patches are aggregated to infer the final image-level label for the WSI. In such methods, accurate patch-level image classification is fundamental to reach the accuracy level of human pathologists.

In the area of cervical histopathology analysis, existing literature~\cite{chankong2014automatic, guo2016nuclei} have studied various supervised learning methods for nuclei-based cervical cancer classification.
Chankong~\textit{et~al.}~\cite{chankong2014automatic} proposed automatic cervical cancer cell segmentation and classification using fuzzy C-means (FCM) clustering and various types of classifiers.
Guo~\textit{et~al.}~\cite{guo2016nuclei} designed hand-crafted nuclei-based features for fusion-based classification on digitized epithelium histopathology slides with linear discriminant analysis (LDA) and support vector machines (SVM) classifier.  While accomplishments have been achieved with fully-supervised learning methods, the training of models require large amounts of expert annotations of cervical histopathology images. Since the annotation process can be expensive, tedious, and time-consuming, it often results in limited or insufficient number of labeled data available for supervised learning models.

\subsection{Conditional Image Synthesis}
Generative adversarial networks (GANs)~\cite{goodfellow2014generative} as an unsupervised learning technique, has enabled a wide variety
of applications including image synthesis, object detection~\cite{li2017perceptual} and image segmentation~\cite{xue2018segan}. Among variants of GANs, conditional GAN (cGAN) generates~\cite{mirza2014conditional, odena2017conditional} more interpretable results with conditional inputs. For instance, images can be generated conditioning on class labels, which enables cGAN to serve as a tool to generate labeled samples for synthetic augmentation. Current state-of-the-art cGAN models often breaks the task into smaller gradual generation or refinement sub-tasks~\cite{zhang2018stackgan++, karras2017progressive}, or employs large scale training~\cite{brock2018large}, which enable them to generate high fidelity images. 
In this work, we use our proposed HistoGAN, which is inspired by state-of-the-art cGANs~\cite{zhang2018stackgan++, zhang2019self, brock2018large}, to generate high-fidelity synthetic images to augment classification model training. To improve the quality of synthetic images and stabilize the training process, our model utilizes numerous techniques including minibatch discrimination~\cite{salimans2016improved}, self attention~\cite{vaswani2017attention}, class conditional batch normalization~\cite{de2017modulating}, and spectral normalization~\cite{miyato2018spectral} following prior art. While generating visually appealing histopathology images, HistoGAN serves as an essential prerequisite for the synthetic data augmentation.

\subsection{Synthetic Data Augmentation}
To better utilize training data and reduce over-fitting during the training process, data augmentation has become a common practice for training deep neural networks. The objective of augmentation is to add to the original training set new samples that follow the original data distribution.  Therefore, a good augmentation scheme should generate samples that follow the original data distribution but are different from those in the original training set. On the other hand, a bad augmentation scheme can generate samples that deviate from the original data distribution thus can mislead training when added to the training set. 

Traditional data augmentation~\cite{wang2017effectiveness} often involves transformations applied directly on original training data, such as cropping, flipping and color jittering. While serving as an implicit regularization, straightforward data augmentation techniques are limited in augmentation diversity. To overcome the limitation of traditional augmentation, several works have been done to improve the effectiveness of data augmentation. Rather than using a pre-defined augmentation policy, Auto Augmentations~\cite{cubuk2019autoaugment, ho2019population} use hyper-parameter searching to automatically find the optimal augmentation policy. 

Another popular trend is to generate synthetic images to increase the amount and diversity of original training data, which we denote as \textit{Synthetic Augmentation}. Along this direction, for natural images, Ratner~\textit{et~al.}~\cite{ratner2017learning} learns data transformation with unlabeled data using GANs. GAGAN~\cite{antoniou2017data} and BAGAN~\cite{mariani2018bagan} uses cGANs~\cite{mirza2014conditional} generated samples to augment the standard classifier in the low-data regime. Compared with works done in the natural image domain, issues related to insufficient and imbalanced data are more prominent in the medical image domain. To mitigate these problems, researchers have been working on synthetic augmentation for medical image recognition tasks. 
Frid-Adar~\textit{et~al.}~\cite{frid2018gan} proposes to use cGAN generated synthetic CT images to improve the performance of CNN in liver lesion classification. Gupta~\textit{et~al.}~\cite{gupta2019generative} synthesizes lesion images from non-lesion ones using CycleGAN~\cite{zhu2017unpaired}. Bowles~\textit{et~al.}~\cite{bowles2018gan} uses GAN derived synthetic images to augment medical image segmentation models. Zhao~\textit{et~al.}~\cite{zhao2018synthesizing} proposes a GAN model for synthesizing retinal images from small sized samples and uses the synthetic images to improve semantic segmentation performance. Mahapatra~\textit{et~al.}~\cite{mahapatra2018efficient} applies a Bayesian neural network (BNN)~\cite{mackay1992practical} to calculate the informativeness of the synthetic images for improved classification and segmentation results. Zhao~\textit{et~al.}~\cite{zhao2019data} uses transformations of labeled images for  one-shot image segmentation.  GAN based synthetic augmentation has achieved promising results, but typically blindly adds synthetic samples to the original data. Few consider how to assure the quality of synthetic images or control the augmentation step after image synthesis.

\subsection{Our Previous Work}
In our recent work~\cite{xue2019synthetic}, we propose a feature based filtering mechanism for synthetic augmentation. While improving classification performance, our previous cGAN generated images are not realistic enough and the work lacks rigorous study of its GAN model training and feature extractor training processes.
In this work, we propose an improved GAN model for histopathology image generation, and develop a more general synthetic augmentation framework by reducing the randomness in GAN model and feature extractor training through MC-sampling and FID score based model selection. Our new contributions and differences from previous work
are summarized as follows:

\begin{itemize}
\setlength{\itemsep}{-0.2em}

\item 
We design and utilize an improved conditional GAN model architecture, namely HistoGAN, with a self-attention module among other techniques to stabilize the training and improve the quality of synthetic images.

\item 
We propose a more general selective synthetic augmentation method which achieves better performances than our previous method.

\item 
We conduct more comprehensive experiments including more ablation study and new results on the PCam dataset.
\end{itemize}

\begin{figure}[t]
\begin{center}
  \includegraphics[width=0.99\linewidth]{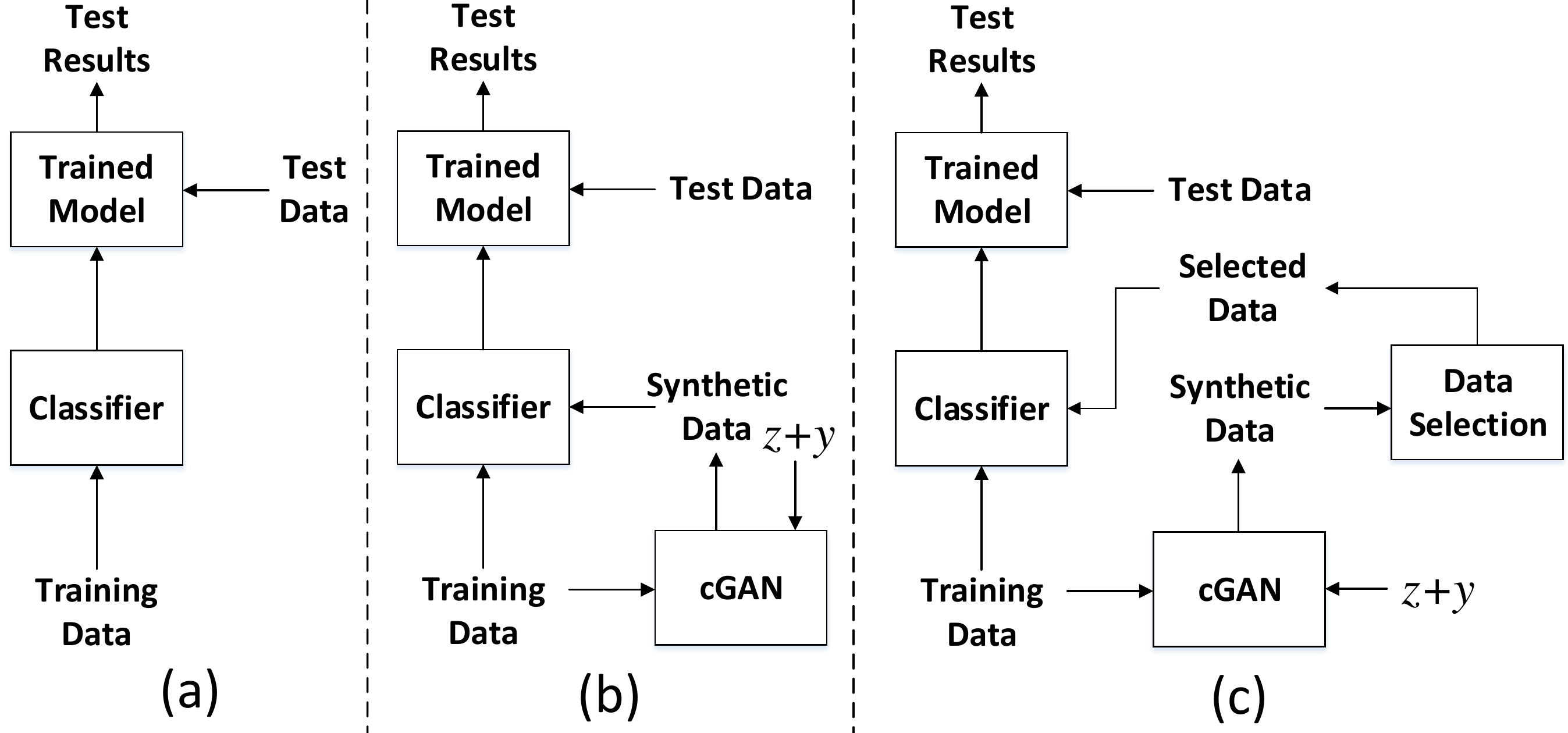}
\end{center}
  \caption{Comparison between different training processes. (a) Traditional training pipeline; (b) Conditional GAN augmented training pipeline; (c) Our proposed selective synthetic augmentation with quality assurance. The input to the cGAN are noise vector $z$ and label condition vector $y$.}
\label{fig:arch_comparison}
\end{figure}

\begin{figure*}[ht]
\begin{center}
  \includegraphics[width=0.95\linewidth]{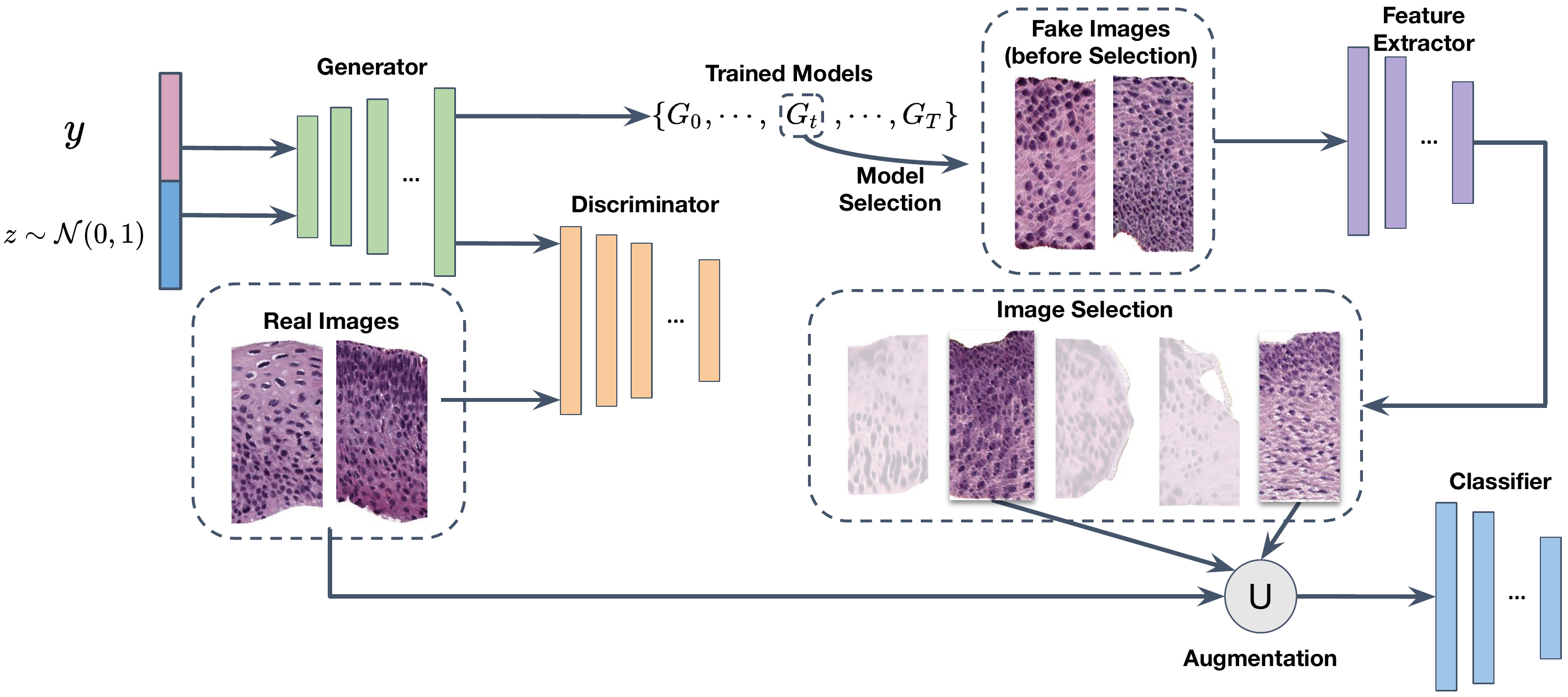}
\end{center}
  \caption{The architecture of the proposed selective synthetic augmentation algorithm. The $\cup$ symbol indicates that the selected synthetic image set is unioned with the original training set to improve classification model training and test performance. }
\label{fig:architecture}
\end{figure*}
\section{Methodology}

In traditional fully-supervised training methods, the model is trained on training images and the inference is done by feeding the test data to the trained model. In previous GAN-based augmentation works~\cite{frid2018gan,madani2018chest}, a GAN model is first trained to generate some synthetic images based on the training data, then the generated images are added to the original training data as a data augmentation strategy. 
However, since the discriminator in GAN only outputs a high level judgement ($0$ or $1$) of the fidelity of generated images, such pipelines cannot guarantee that the generated data contain meaningful features which contribute to improving classification model training. To tackle this issue, we propose a selective synthetic augmentation algorithm to evaluate the quality and fidelity of synthetic images and select only those samples with high-confidence in label correctness and real-image likeness to be added to the training set. 
The comparison between different training procedures is illustrated in Fig.~\ref{fig:arch_comparison}.

An overall illustration of our proposed selective synthetic augmentation method can be found in Fig.~\ref{fig:architecture}. We first train a conditional GAN model based on the labeled training images. The optimal model weights is selected based on the smoothed FID score~\cite{heusel2017gans}. A pool of synthetic images are then generated using the selected model. All images are then passed into the image selection module to filter out the ones that fail to contribute sufficient amount of meaningful information. After image selection, a classification model is trained with both original and synthetic training data. Trained classification models can then be used for inference on test data. More details are introduced in the following subsections. 

\subsection{HistoGAN Model} 
In this section, we introduce our proposed HistoGAN architecture and how to select the best model with highest synthetic image quality from a set of trained models.

\subsubsection{Model Architecture} \label{sec:architecture}
The conventional cGANs~\cite{mirza2014conditional} have an objective function defined as:
\begin{multline}
\min_{\theta_G} \max_{\theta_D} \mathcal{L}_{\text{cGAN}} = \mathbb{E}_{x\sim P_\text{data}}[\log D(x,y)] + \\ \mathbb{E}_{z\sim \mathcal{N}}[\log (1 - D(G(z,y)))] \enspace .\label{Eq:cGAN}
\end{multline}
In the equation above, $x$ represents the real data from an unknown image distribution $P_\text{data}$ and $y$ is the conditional label (\emph{e.g.}, CIN grades). $z$ is a random vector for the generator $G$, drawn from a standard normal distribution $\mathcal{N}(0,1)$. During the training, $G$ and $D$ are alternatively optimized to compete with each other.

\begin{figure*}[ht]
\begin{center}
  \includegraphics[width=0.95\linewidth]{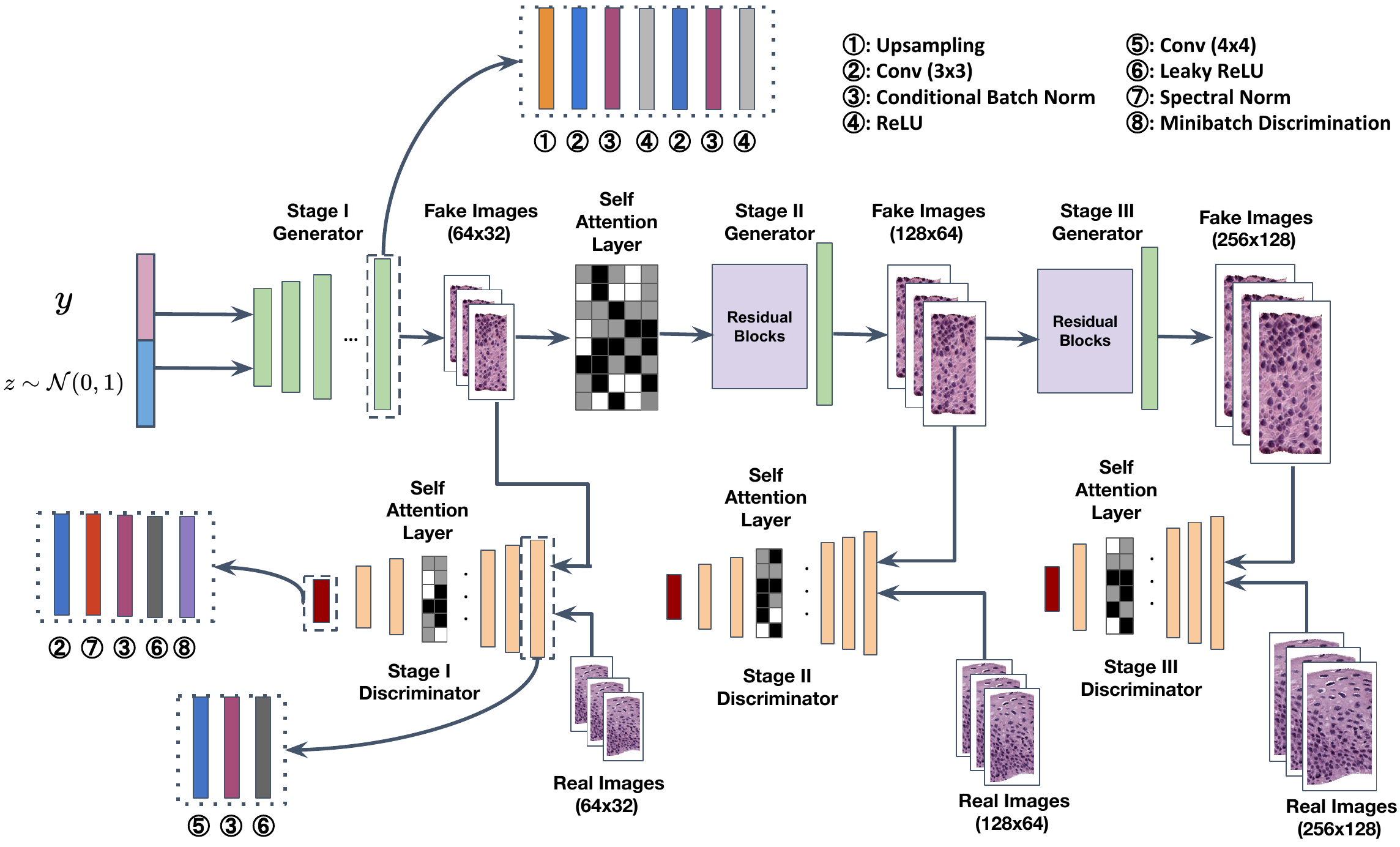}
\end{center}
  \caption{The architecture of a 3-stage HistoGAN for cervical epithelium synthesis. The number of stages can be adjusted according to the desired final image resolution. Detailed features such as cytoplasm texture and nuclei shapes get progressively refined in synthetic images of higher resolution from stage I to III. The self attention layer is applied after stage I generator where the sketch outline and rough pattern of images are shaping up. Self attention layers are also incorporated in discriminators at all stages to further enforce the consistency of focused local regions more accurately. Conditional batch normalization~\cite{de2017modulating} is used after convolutional layers for flexibly modulating convolutional feature maps.}
\label{fig:net}
\end{figure*}

Since there is no existing cGAN framework specifically designed for histopathology image synthesis, we choose to design a new model, \textit{HistoGAN}, based on previous state-of-the-art conditional GAN models and techniques~\cite{zhang2018stackgan++,brock2018large,zhang2019self}. We aim to generate synthetic images in a coarse-to-fine fashion through multiple stages, where details of images are gradually refined to guarantee the fidelity. The training procedure of HistoGAN is similar to Eq.~\ref{Eq:cGAN}. The generator of the first stage takes a random noise vector and class label as input, and the generator of remaining stages will take the output of the previous stage as input instead of random noise. To
increase diversity among the generated examples and mitigate the issue of mode collapse indicated by the high homogeneity of the synthetic image pool, we incorporate the minibatch discrimination module~\cite{salimans2016improved} into our discriminator.
Following state-of-the-art works in conditional image synthesis~\cite{zhang2019self,brock2018large}, class conditional batch normalization is used in both generators and discriminators to enhance the learning effectiveness of the inter class feature discrepancy. And spectral normalization~\cite{miyato2018spectral} is utilized in discriminators of all stages to further improve model performance.

To better capture the distribution of nucleus density and color changes in histopathology images of different classes, we leverage self attention~\cite{vaswani2017attention,zhang2019self} at early stages of generation and throughout all stages in the discrimination process. 
The application of self attention mechanism enables both generator and discriminator to better learn the dependencies between spatial regions by looking at the relationship between one pixel and all other positions in the same image. Similar to~\cite{zhang2019self}, the image features from the previous hidden layer $x$ are first transformed into two feature spaces $q, k$ as query and key in self attention~\cite{vaswani2017attention}
to calculate the attention map. Let $q(x) = W_qx$ and $k(x) =
W_kx$, the attention map over the $i$th location when synthesizing the $j$th region is

\begin{equation}
\alpha_{j, i}=\frac{\exp \left(s_{j i}\right)}{\sum_{i=1}^{N} \exp \left(s_{j i}\right)}, \text { where } s_{j i}=\boldsymbol{q}\left(\boldsymbol{x}_{i}\right)^{T} \boldsymbol{k}\left(\boldsymbol{x}_{j}\right)\enspace.
\label{eq:sa1}
\end{equation}

The output of the self attention of the $j$th region $o_j$ is calculated by applying attention weight over the value $v$ as
\begin{equation}
\boldsymbol{o}_{j}=\sum_{i=1}^{N} \alpha_{j, i} \boldsymbol{v}\left(\boldsymbol{x}_{i}\right), \text { where } \boldsymbol{v}\left(\boldsymbol{x}_{i}\right)=\boldsymbol{W}_{v} \boldsymbol{x}_{i}\enspace.
\label{eq:sa2}
\end{equation}
In all transformation matrices $W_q, W_k$, and $W_v$, weight matrices are implemented as $1\times1$ convolutions. Compared with the StackGAN model implemented in our previous work~\cite{xue2019synthetic}, our HistoGAN generates more realistic image patches which also benefits the following synthetic augmentation step. An example of cGAN result comparison is shown in Fig.~\ref{fig:cervical}.

\subsubsection{Model Selection}\label{model_selection}
During training of the HistoGAN model, the model weights vary from epoch to epoch.  A challenge is to determine which model weights gives rise to better synthetic image quality. For natural image synthesis tasks, Inception Score~\cite{salimans2016improved} and Fréchet Inception Distance (FID)~\cite{heusel2017gans} score are two commonly used metrics. The calculation of these two metrics rely on the pre-trained Inception V3~\cite{szegedy2016rethinking} model trained on ImageNet~\cite{deng2009imagenet}. However, since the distribution of natural images and that of medical images such as cervical histopathology images can be quite different, we can not directly use the aforementioned two scores for evaluating our HistoGAN model. 
Instead, we follow the calculation of the original FID score while replacing the Inception V3 model pre-trained on ImageNet with a ResNet34~\cite{he2016deep} model pre-trained on the cervical histopathology dataset.

\begin{figure*}[t]
\begin{center}
  \includegraphics[width=0.99\linewidth]{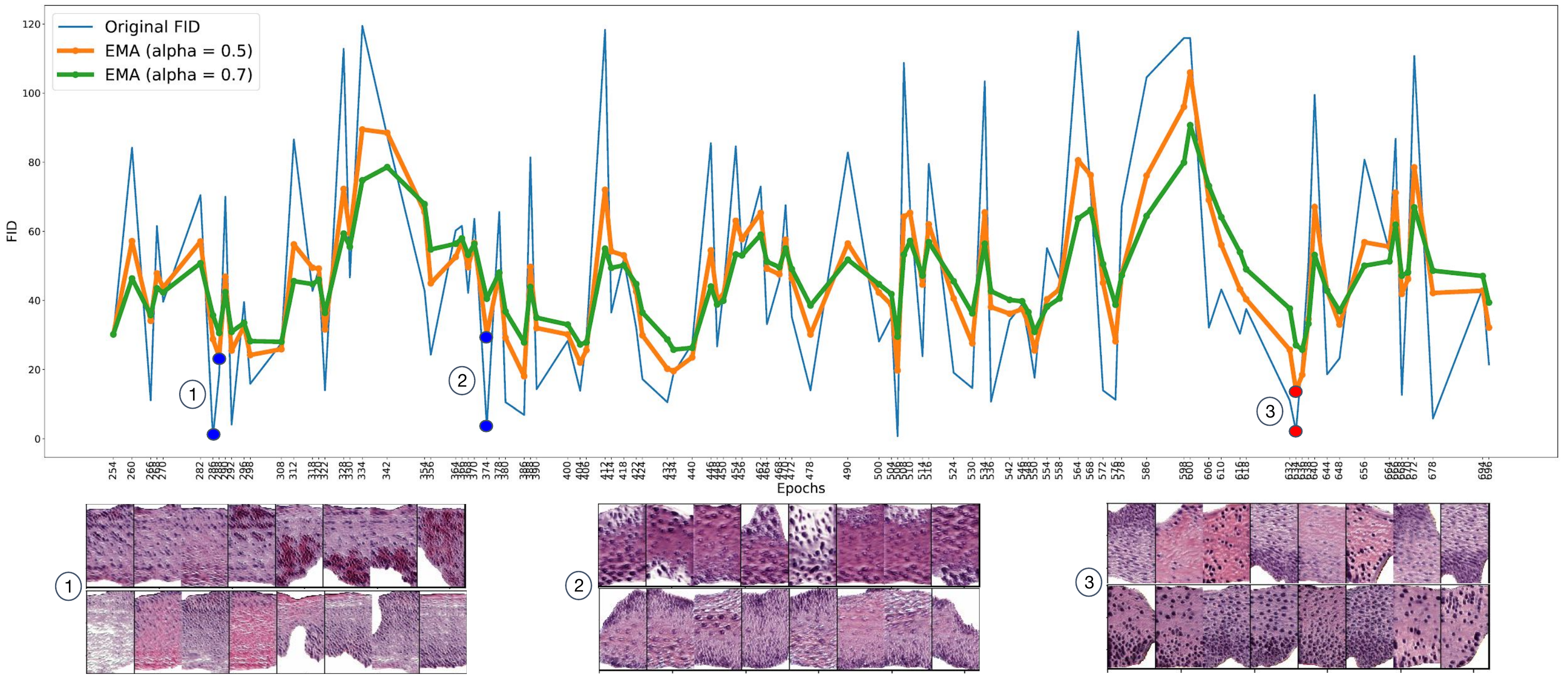}
\end{center}
  \caption{FID scores (pre-trained ResNet34) of HistoGAN models saved after different number of epochs of training. Scores are smoothed with varying EMA parameter $\alpha$.}
\label{fig:fid}
\end{figure*}

To compare the trained models after running different numbers of epochs, we save the HistoGAN model after each epoch of training. To estimate the performance of each saved model, we calculate the FID score between the feature vectors of real and generated images extracted from the pre-trained ResNet34 model as follows:

\begin{small}
\begin{multline}
     d(x, \tilde{x}) = \left|\left| \mu_{x \sim P_\text{data}}\phi(x) - \mu_{\tilde{x} \sim P_G}\phi(\tilde{x}) \right|\right|_{2}^{2} + \\ 
     \mathbf{Tr} \left (\Sigma_{x \sim P_\text{data}}\phi(x) + \Sigma_{\tilde{x} \sim P_G}\phi(\tilde{x}) - 2 \left (\Sigma_{x \sim P_\text{data}}\phi(x) \Sigma_{\tilde{x} \sim P_G}\phi(\tilde{x}) \right )^{\frac{1}{2}} \right ) \enspace ,
\end{multline}
\end{small}

 where $\tilde{x}$ represents synthetic images generated by the saved HistoGAN model being evaluated, and $\phi$ denotes the features extracted from intermediate layers of the pre-trained ResNet34 model. Assume feature vectors follow a multivariate Gaussian distribution, the mean and covariance are estimated for the real and fake data~\cite{borji2019pros} for fréchet distance calculation to measure the visual quality of generated images. Smaller FID scores indicate better visual quality.
Although the FID score itself cannot guarantee agreement with human judgment, trends of FID often provide a reliable estimation of the quality of a GAN model. As we can observe from Fig.~\ref{fig:fid}, due to the instability in GAN training, the FID scores of each saved epoch fluctuate constantly and fail to provide a distinguishable pattern. Based on the unaltered FID scores, one should choose the model saved at epoch 286 or epoch 374. However, one can see that images generated by these chosen models are not satisfactory as in Fig.~\ref{fig:fid}. To get a robust estimation of model quality and mitigate the effect caused by outliers, we apply the Exponential Moving Average (EMA)~\cite{hunter1986exponentially} algorithm to smooth the curve of original FID score. With smoothing, the FID score at time $t$ is:

\begin{equation}
    \hat{d} = \left\{\begin{matrix} \text{d}_t,& \enspace t=1
\\ \alpha \hat{d}_{t-1} + (1-\alpha)\text{d}_t,& \enspace t>1 \label{Eq:smoothFID}
\end{matrix}\right.
\end{equation}

We monitor the training process with the smoothed FID. As shown in Fig.~\ref{fig:fid}, different values of $\alpha$ lead to different levels of smoothing in FID and we observed that the chosen model associated with the lowest smoothed FID score has better image quality than the model chosen using the lowest original FID score. In our experiments, we set $\alpha$ to $0.5$ for a medium level of smoothing. One can see that, after smoothing with the EMA algorithm, the minimum in smoothed FID score is reached at epoch 634, which is the GAN model we chose for the follow-on synthetic data augmentation.

\subsection{Image Selection}\label{img_selection}
\begin{figure}[t]
\begin{center}
  \includegraphics[width=0.95\linewidth]{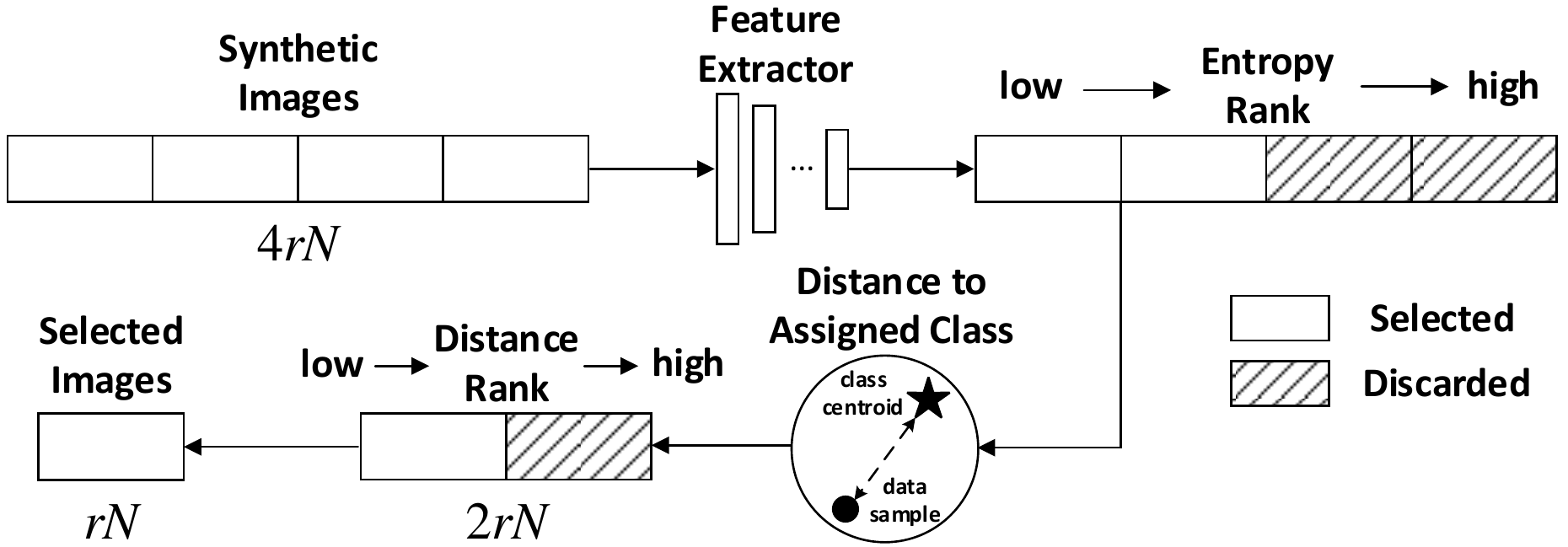}
\end{center}
  \caption{Illustration of the image selection process. $r$ and $N$ represent the augmentation ratio and the number of original training data. The same feature extractor runs multiple times through MC-dropout for both entropy and class centroid distance calculations to increase robustness.}
\label{fig:img_selection}
\end{figure}

Given a trained cGAN model, one can sample infinite number of noise-vector inputs from the Gaussian distribution and generate infinite number of synthetic images. While a good cGAN model can generate images that look real, there are no guarantee that those images would be good to be used for augmenting the original training set in visual recognition tasks. In current GAN-based data augmentation methods, with different data augmentation ratio, different number of generated images are added to the training set. However, the effectiveness of such augmentation pipeline is heavily affected by the varying quality of synthetic images as well as the diversity of the images. To reduce the randomness in the synthetic augmentation process and selectively add in new images, we break the whole process into two steps: find samples that can be confidently classified into certain classes thus containing enough diagnosable features; then find samples whose features are within a certain neighborhood of class centroids in the feature space to assure matching between the synthetic image and its assigned label. Such steps are done with a pre-trained feature extractor to calculate centroids for real samples and extract features for fake samples. Considering that a single feature extractor cannot provide robust feature extraction results, we use a feature extractor with Monte Carlo dropout (MC-dropout)~\cite{gal2016dropout} and take the expectation value of multiple samplings to reduce the uncertainty of feature extraction. A depiction of our proposed selective synthetic augmentation algorithm is shown in Fig.~\ref{fig:img_selection} and a detailed description is given in Algorithm~\ref{algorithm}.

The first step of selection is based on label certainty of a sample. In traditional machine learning systems, real samples that lie near the decision boundary are often assumed to contain more important features for classification purposes. However, as we conducted experiments to select good synthetic images, one interesting finding is that selecting the fake samples with more certain labels gives better classification performance than selecting those with less certain labels. 
This may be due to the cGAN model being imperfect and conditionally-generated fake examples with less label certainty being more likely to deviate from the real data distribution. 
In our algorithm, we evaluate the label certainty of a fake example by calculating the entropy score of its predicted class probabilities. 
If the feature extractor is certain that a sample can be classified into a certain class, the entropy score would be low.  We rank the entropy scores of all generated images in ascending order and choose the first half of images with lower entropy. 
The necessity of this entropy-based selection is proved by experiments on different datasets, which will later be discussed in Section~\ref{Experiments}.

\begin{algorithm}[!t]
\begin{algorithmic}
    \caption{Selective Synthetic Augmentation}\label{algorithm} 
    \STATE \textbf{Input}: a set of trained HistoGAN models \{$G_t$\}, number of classes $\mathcal{C}$, augmentation ratio $r$, number of original training samples $N = \sum_{i=1}^{\mathcal{C}} N_i$.
    \STATE \textbf{Output}: selected synthetic samples $\mathcal{X}$ with $|\mathcal{X}|=rN$.
    \STATE \textbf{Initialization}: $\mathcal{X}_1 = \emptyset$,
    $\hat{t} = \arg \min (\hat{d_t})$,
    $G_{\hat{t}}$ generated samples $\mathcal{X}_{0}$ = \{$x_j^i: i \leq \mathcal{C}, j \leq 4rN_i$ \},
    entropy $\mathcal{E}^i = \{e_j^i: e_j^i = - \sum p_j^i \log p_j^i, i \leq \mathcal{C}, j \leq 4rN_i \}$. \\
    \FOR {$x_j^i \in \mathcal{X}_0$}
    \IF {$e_j^i < \text{Median} (\mathcal{E}^i)$}
    \STATE $\mathcal{X}_1 = \mathcal{X}_1 \cup \{x_j^i\}$ 
    \ENDIF
    \ENDFOR \\
    class centroid distance $\mathcal{D}^i = \{d_j^i: d_j^i = D_f (x_j^i, c_i)\}$. \\
    \FOR {$x_j^i \in \mathcal{X}_1$} 
    \STATE $d_j^i = D_f \{x_j^i, c_i\}$ 
    \IF{$d_j^i < \text{Median} (\mathcal{D}^i)$}
    \STATE $\mathcal{X} = \mathcal{X} \cup \{x_j^i\}$ 
    \ENDIF
    \ENDFOR
\end{algorithmic}
\end{algorithm}

After the entropy selection step, we further select synthetic images based on their distance to class centroids in the feature space. In this second step of selection, all remaining samples that have passed the entropy-based selection will have their feature distances to their class centroids calculated. All distances will be sorted in ascending order and the first half of these samples with smaller distances will be kept.
The motivation behind ranking samples based on their feature distance to class centroids is to help filter out samples whose assigned labels (i.e. the conditional labels used by the cGAN 
model to generate them) do not match their classified labels in feature space so that only samples that confidently match with their assigned labels are selected and added to the training set.
In our implementation, instead of using a single run of the feature extractor to extract features, we run the feature extractor multiple times with MC-sampling and then calculate feature distances based on the average feature distance from the multiple runs. 
Similar to~\cite{xue2019synthetic}, the feature distance between image $x$ and centroid $c$ is defined as 

\begin{equation}
D_{f}(x,c_i) = \frac{1}{K}\sum_{k}\sum_{l} \frac{1}{H_lW_l} \left|\left|\hat{\phi_l^k}(x) - \hat{\phi_l^k}(c_i)\right|\right|_{2}^{2} \enspace ,\label{Eq:feature}
\end{equation}

\noindent where $\hat{\phi_l^k}$ is the unit-normalized activation in
the channel dimension $A_l$ of the $l$th layer of the $k$-th MC-sampling feature extraction network with shape $H_l \times W_l$. We denote the total sampling time as $K$. $D_{f}(x,c_i)$ can be regarded as an estimated cosine distance between sample and $i$-th centroid in the feature space. 

The centroid $c$ is calculated as the average feature of all labeled training images in the same class. For class $i$, its centroid $c_i$ is represented by
\begin{equation}
c_i = \left[\frac{1}{N_i}\sum_{j=1}^{N_i}\phi_1(x_j),...,\frac{1}{N_i}\sum_{j=1}^{N_i}\phi_L(x_j)\right] \enspace ,\label{Eq:centroid}
\end{equation}
\noindent where $N_i$ denotes the number of training samples in $i$th class and $x_j$ is the $j$th training sample. Similar to Eq.~\ref{Eq:feature}, $\phi_l$ is the activation extracted from the $l$th layer of the feature extraction network. $L$ is the total number of layers utilized in the feature distance selection. $c_i$ is retained by one time MC-sampling and fixed during the distance calculation. 
 
 In conclusion, given augmentation ratio $r$, we first generate $4rN_i$ images for each class $i$, then select $rN_i$ images according to the two-step selection process described above. Regarding the choice of $r$, we provide an
 ablation study in Section~\ref{sec:results}. 
 
\section{Experiments}\label{Experiments}

\begin{figure*}[!ht]
\begin{center}
  \includegraphics[width=0.98\linewidth]{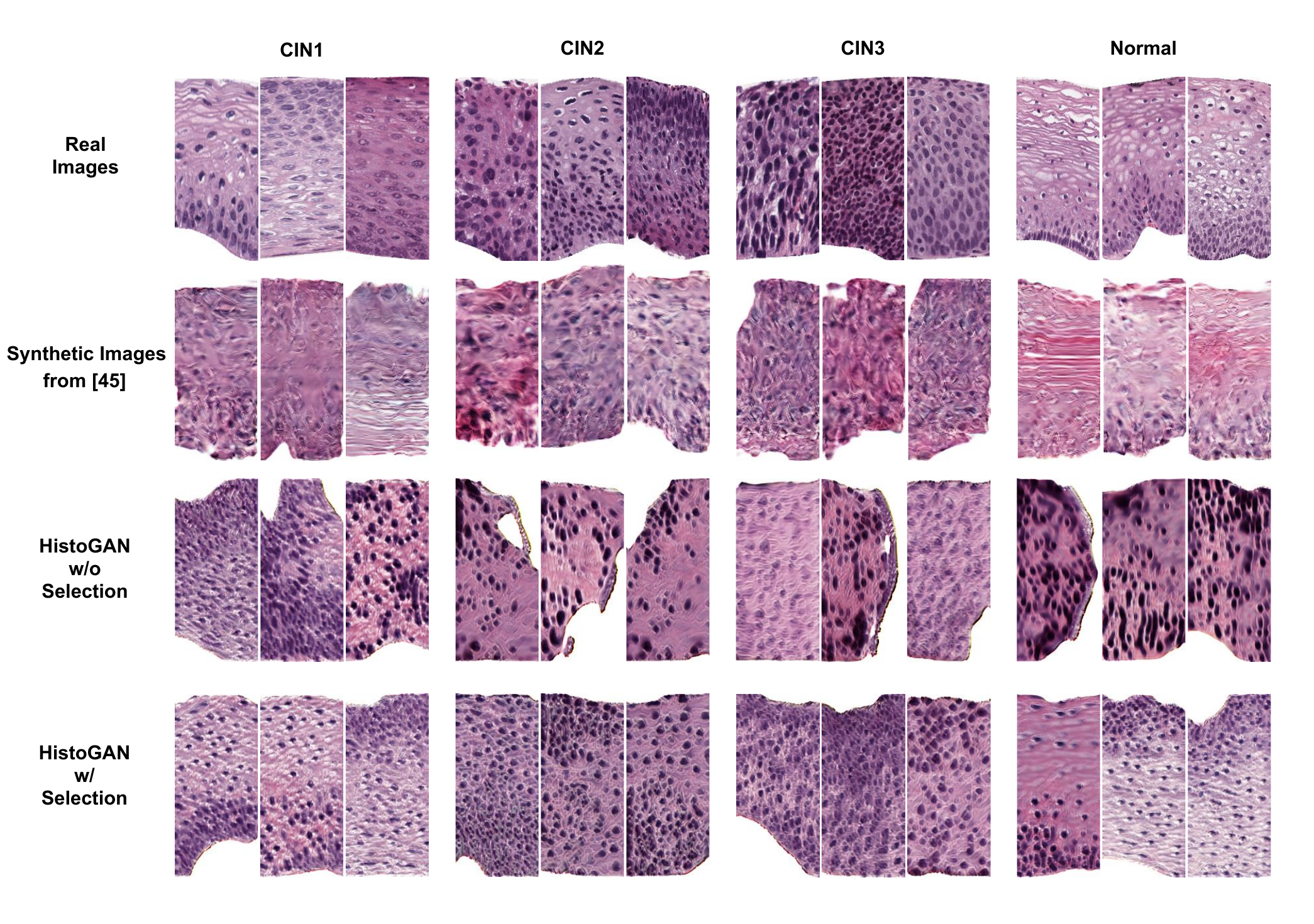}
\end{center}
  \caption{Examples of real images, synthetic images generated from~\cite{xue2019synthetic}, and images generated by our HistoGAN model trained on cervical histopathology dataset before and after selection. Our HistoGAN generates realistic images with clearly better visual quality than those by ~\cite{xue2019synthetic}. Zoom in for better view.}
\label{fig:cervical}
\end{figure*}

\begin{figure*}[!ht]
\begin{center}
  \includegraphics[width=0.98\linewidth]{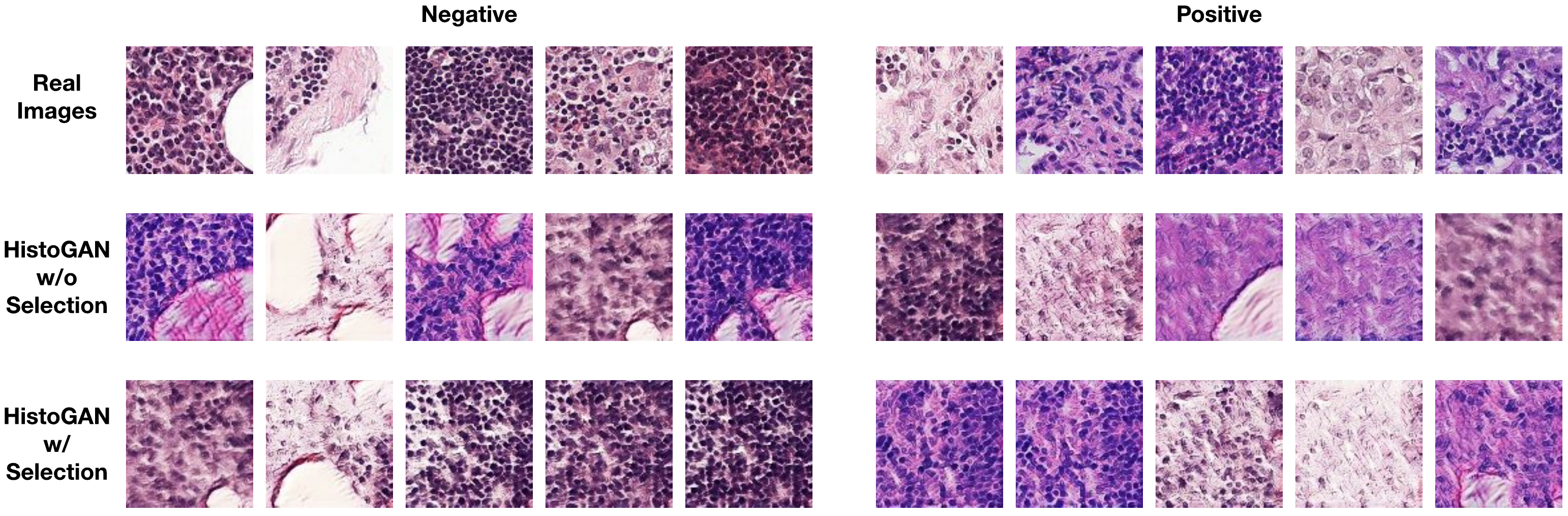}
\end{center}
  \caption{Examples of real and synthetic images generated by HistoGAN trained on 10\% of PCam dataset.}
\label{fig:pcam}
\end{figure*}

\begin{figure*}[!ht]
\begin{center}
  \includegraphics[width=0.98\linewidth]{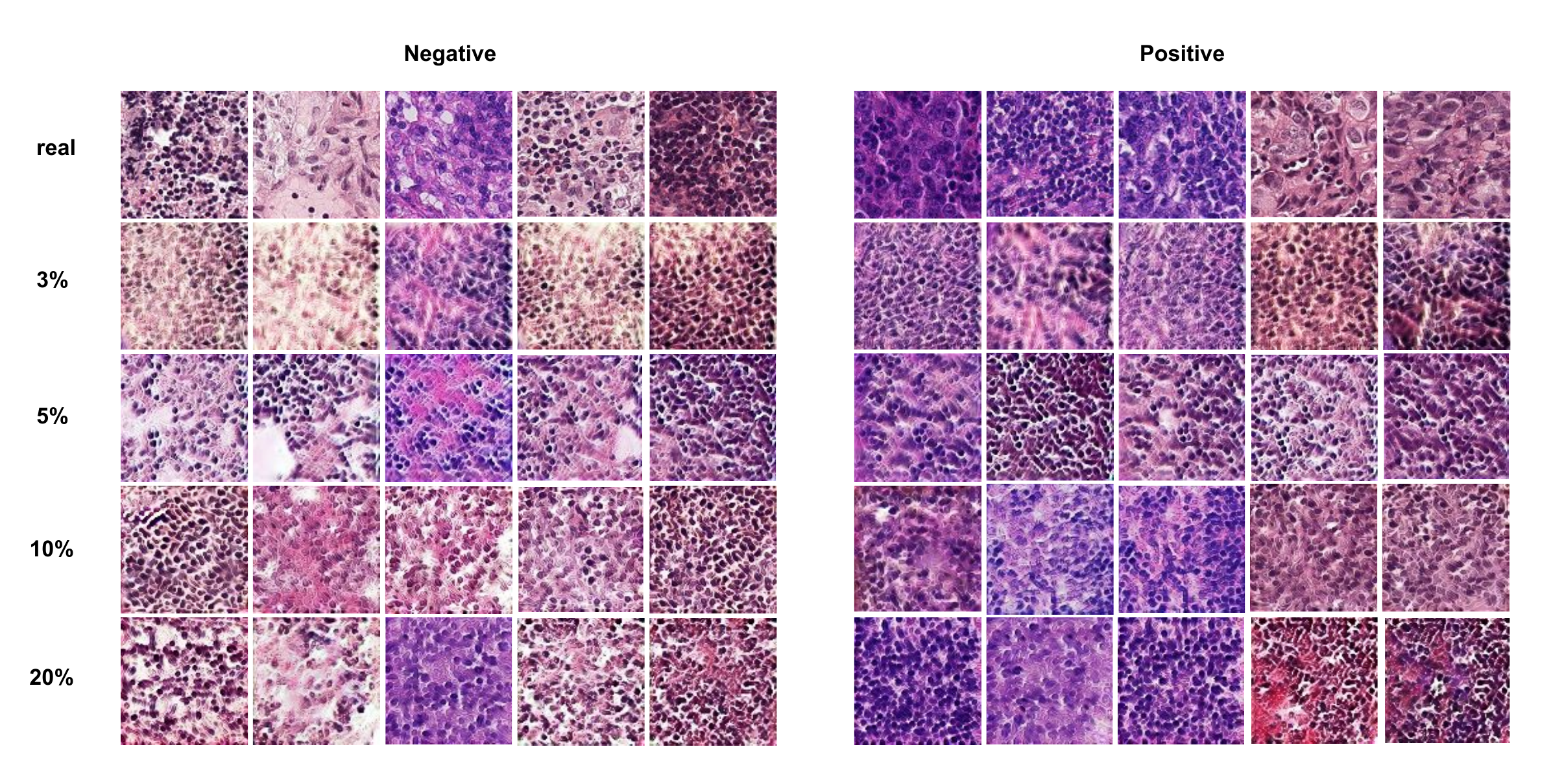}
\end{center}
  \caption{Examples of real and synthetic images generated by HistoGAN trained on 3\%, 5\%, 10\% and 20\% of PCam dataset. All generated images are chosen from the pool after applying our proposed image selection method. Zoom in for better view.}
\label{fig:pcam_full}
\end{figure*}

\begin{figure*}[!ht]
\begin{center}
  \includegraphics[width=0.98\linewidth]{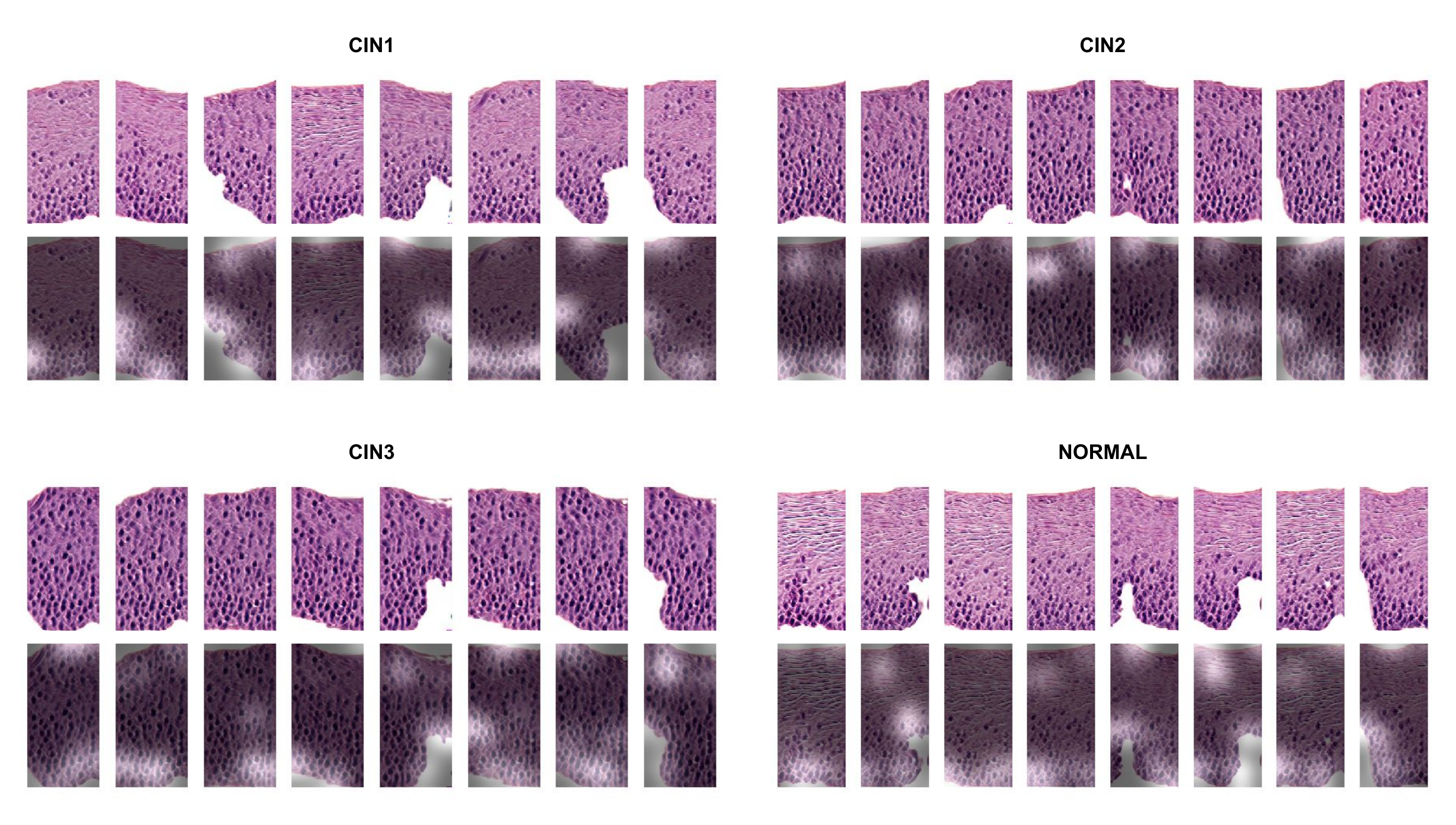}
\end{center}
  \caption{The attention map extracted from the self attention layer applied after stage I generator as illustrated in Figure \ref{fig:architecture}. The first row shows the synthetic images generated by our proposed HistoGAN model; the second row gives the most attended regions of each image during the GAN training phase by overlaying the attention map on top of the original image. Higher attention scores correspond to the highlighted areas where distinguishing patterns like cell crowding and nuclei distribution are highlighted. It demonstrates the effectiveness of the attention mechanism incorporated in our HistoGAN model.}
\label{fig:self_attn_map}
\end{figure*}

\subsection{Datasets}
The first dataset contains labeled cervical histopathology images collected from a collaborating health sciences center. All images are annotated by the same pathologist. The data processing follows~\cite{xue2019synthetic}, and results in patches with a unified size of $256 \times 128$ pixels. Compared with the dataset used in~\cite{xue2019synthetic}, we include more data for more comprehensive experiments. In total, there are $1,284$ Normal, $410$ CIN1, $481$ CIN2, $472$ CIN3 patches. Examples of the images can be found in the first row of Fig.~\ref{fig:cervical}. 
We randomly split the dataset, by patients, into training, validation, and testing sets, with ratio 7:1:2 and keep the ratio of image classes almost the same among different sets. All evaluations and comparisons reported in this section are carried out on the test set.

To further prove the generality of our proposed method, we also conduct experiments on the public PatchCamelyon (PCam) benchmark~\cite{veeling2018rotation}. PCam consists of $327,680$ color patches extracted from histopathologic scans of lymph node sections with unified size of $96 \times 96$ pixels. The PCam dataset is split into 75\%:12.5\%:12.5\% of training, validation, and testing
sets, selected using a hard-negative mining regime. Each image is annotated with a binary label indicating presence of metastatic tissue. To mimic the situation where only a limited amount of training data is available, we use randomly selected $10\%$ of the training set, which has 32,768 patches, to train our proposed HistoGAN model and the baseline classifier. Trained models are evaluated on the full test set.

\begin{figure*}[!ht]
\begin{center}
  \includegraphics[width=0.99\linewidth]{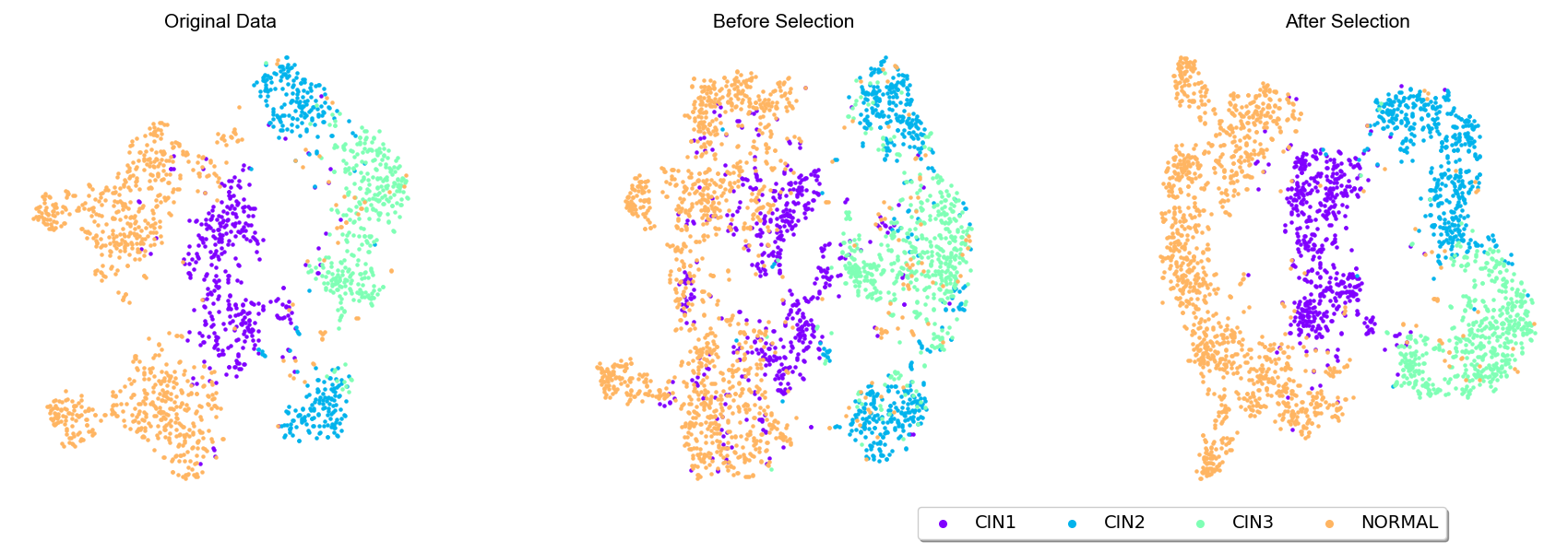}
\end{center}
  \caption{t-SNE of the original and augmented cervical histopathology training set before and after image selection. The augmented training data after selection clearly have more distinguishable features than the ones without selection.}
\label{fig:tsne_cervical}
\end{figure*}

\begin{figure*}[!ht]
\begin{center}
  \includegraphics[width=0.99\linewidth]{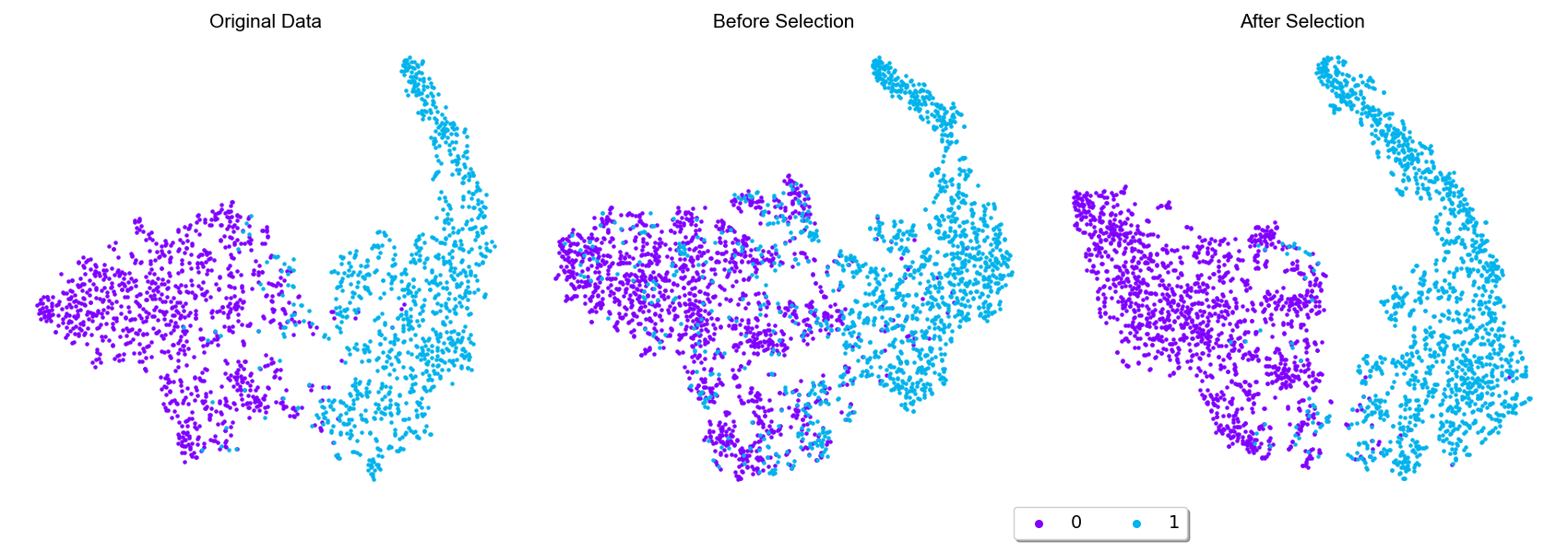}
\end{center}
  \caption{t-SNE of the original and augmented PCam histopathology training set before and after image selection. While data augmentation without image selection increases the number of training samples, the original data distribution is distorted. After image selection, the original data distribution is recovered along with more number of data points. }
\label{fig:tsne_pcam}
\end{figure*}

\subsection{Implementation Details}

\subsubsection{HistoGAN Implementation}

The proposed HistoGAN model is trained in parallel on 4 NVIDIA TITAN Xp GPUs, each with 11G of RAM. We train HistoGAN with WGAN-GP~\cite{gulrajani2017improved} loss on the discriminators at all stages. Based on different sizes of images in the training set, we construct a 3-stage HistoGAN for cervical histopathology images and a 2-stage HistoGAN for the PCam lymph node histopathology images.

The input of the generator at the first stage is the concatenation of random noise and class label (\textit{e.g.}, CIN1-3, Normal) that are first one-hot encoded and then embedded by a transposed convolution layer. The first stage generator consists of $4$ up-sampling blocks with $3\times3$ conv kernels. Each block contains an upsample layer with bilinear interpolation followed by a combination of a convolutional layer with $3\times3$ kernel size. The output then goes through a conditional batch normalization \cite{de2017modulating} layer to modulate convolutional feature maps based on the corresponding assigned labels of the images generated. Blocks of the same architecture but different in and out channels are employed in generators of the next stages respectively, after a set of residual blocks. 

Considering the future stages are learning the features from a more granularized level based on the output of the first stage, we employ self attention right after the first stage to facilitate the learning and focus on the desired features that are decisive for classification. Next, together with the real images from the original dataset with the same resolution, synthetic images of each scale are fed into corresponding stages of discriminators. 
Inside each discriminator, the main structure contains several down-sampling layers with $4\times4$ conv kernels. Similar to the aforementioned blocks in the generator, class-conditional batch normalization are used after each convolutional layer to embed more class specific information. The down-sampling layers are followed by a $3\times3$ conv layer, a spectral normalization layer, a batch normalization layer, a Leaky ReLU activation layer, a minibatch discrimination~\cite{salimans2016improved} block for preventing mode collapse during GAN training, and a fully connected layer for the final output.

Regarding the hyperparameters, the HistoGAN model used for generating cervical histopathology and PCam images are trained with batch size set to $64$ for the cervical and $256$ for the PCam dataset for $1000$ training epochs with fixed learning rate $2e-4$. The parameter $\delta$ for WGAN-GP loss is set to $50$.

\subsubsection{Model and Image Selection Framework}

In the next step, GAN models at each epoch are saved after the $100^{th}$ epoch for model selection.
For reasons mentioned in Section~\ref{model_selection}, the feature extractor used for FID score calculation is the same as our baseline classifier (ResNet34), followed by EMA-based smoothing to accentuate the pattern of synthetic image quality trend during the GAN training process. The optimal GAN model weights selected for further stages of our purposed sample selection corresponds to the epoch with the lowest adjusted FID score. Next, we generate $4 rN_i$ synthetic images for each class $i$ with the chosen GAN, on which the same feature extractor is run for 5 times in order to extract the predicted probability from the softmax layer for entropy calculation, and also extract feature vectors after each residual block to obtain distance to centroids of ground truth. A dropout layer of rate $0.5$ is inserted before the last residual block right above the fully-connected layer of the feature extractor (ResNet34) for Monte Carlo sampling. Then the generated images are ranked based on the mean of entropy across 5 runs in ascending order, of which half images in each class are kept. The selected pool of synthetic images are further ranked based on the mean of cosine distance to the centroid that corresponds to the assigned label of each image over 5 runs also in ascending order. Similarly, half are filtered out, leaving the rest for the final augmentation.

\subsection{Results Analysis} \label{sec:results}
{
\subsubsection{Evaluation by Expert Pathologists}
To evaluate the quality of images generated by the proposed HistoGAN and validate the effectiveness of the selective synthetic augmentation method, we invited two pathologists to conduct expert evaluation on the cervical histopathology dataset. To prepare for the pathologist evaluation, we randomly chose 100 synthetic images where half of them are before selection and the other half are after selection. These images are then divided into 10 groups.  Within each group of 10 images, there are two subgroups of 5 images where one subgroup is from the before-selection set and the other one is from the after-selection set. The 10 groups of images were then presented to the two pathologists who evaluated their quality independently.  For each group, a pathologist was asked to choose one subgroup that has better quality, without knowing which subgroup corresponds to the one after selection; if the two subgroups were considered to have similar quality, the pathologist chose a tie.  After the pathologists completed their evaluation, we compared their selected subgroups with the ground truth about which subgroups are from the after-selection image set. The comparison result shows that the two pathologists were able to differentiate before-selection subgroups from after-selection subgroups with high consistency: among the 10 groups, they chose the after-selection subgroup as having better quality 7 times, they chose a tie 2 times, and only once they chose the before-selection subgroup as having better quality. This evaluation result demonstrates that our image selection method is highly effective, since the expert pathologists consistently chose the after-selection images as having better quality. 

Besides the group-level evaluation of our image selection method, the two pathologists also assessed the quality and realism of the individual synthetic images. They highlighted some realistic characteristics of the synthesized images, such as correct orientation, cell polarity, clear borders, and correct color of the cytoplasm. They also pointed out some unrealistic characteristics that repeatedly appeared in the generated image, such as smudged chromatin, missing nuclear details for large dark nuclei, and incorrect texture of large sheets of keratin. Despite the unrealistic aspects that they saw in the images, the pathologists actually view most of the images as containing meaningful features that make the images diagnosable. We are encouraged by these findings and plan to incorporate such expert knowledge in our future work to further improve our image synthesis model.
}

\subsubsection{Qualitative Evaluation}
The image synthesis results for cervical and lymph node datasets are demonstrated in Fig.~\ref{fig:cervical} and Fig.~\ref{fig:pcam}, respectively. In Fig.~\ref{fig:cervical}, we also show a comparison of synthetic images generated by our previous work \cite{xue2019synthetic} and by our proposed HistoGAN in this work. In both datasets, as we have already achieved promising image generation results, determining whether those samples can be used for data augmentation or not cannot be easily done by human observations. However, the discrepancy between images with and without selection is much more prominent in the feature space. In order to visualize such differences, after training a baseline ResNet34 classifier with the original training data, we use the pre-trained ResNet34 model as the feature extractor to extract features from the last convolutional layer in the ResNet model. We explore the distribution of training samples, including both original images and synthetic images, in the feature space using t-SNE~\cite{maaten2008visualizing}. In Fig.~\ref{fig:tsne_cervical}, without image selection, samples from different classes are
entangled together, introducing obscuring noise that disrupts the data distribution that real data presents. On the contrary, selected images have clearly more distinguishable features and can potentially help with improving the classification model performance. Similar phenomenon is also observed with more noticeable pattern in Fig.~\ref{fig:tsne_pcam}: while data augmentation without image selection increases the number of training samples, the original data distribution is distorted. After image selection, the original data distribution is recovered along with more number of data points.

The self attention mechanism (Sec. \ref{sec:architecture}) is a core improvement of our proposed HistoGAN model in this work as compared to the GAN model used in our previous work \cite{xue2019synthetic}.  In order to examine the role of self attention, we visualize the conditional attention maps for images from different classes in Fig. \ref{fig:self_attn_map}. From the figure, one can see that HistoGAN with self attention successfully learns meaningful features by attending to important areas containing patterns most useful in distinguishing images of different disease grades.

\begin{table*}[t]
\begin{center}
\begin{tabular}{|l|c|c|c|c|}
\hline
{} &         Accuracy &              AUC &      Sensitivity &      Specificity \\
\hline
Baseline Model~\cite{he2016deep}        &  0.754 $\pm$ 0.012 &  0.836 $\pm$ 0.008 &  0.589 $\pm$ 0.017 &  0.892 $\pm$ 0.005 \\
\hline
\enspace + Traditional Augmentation &  0.766 $\pm$ 0.013 &  0.844 $\pm$ 0.009 &  0.623 $\pm$ 0.029 &  0.891 $\pm$ 0.006 \\
\hline
\enspace + GAN Augmentation, r=0.5      &  0.787 $\pm$ 0.005 &  0.858 $\pm$ 0.003 &  0.690 $\pm$ 0.014 &  0.909 $\pm$ 0.003 \\
\hline
\enspace + Single Filtering~\cite{xue2019synthetic}$^{*}$, r=0.5 &  0.808 $\pm$ 0.005 &  0.872 $\pm$ 0.004 &  0.639 $\pm$ 0.015 &  0.912 $\pm$ 0.006 \\
\hline
\enspace + Selective Augmentation, r=0.5   &  \textbf{0.821 $\pm$ 0.011} &  \textbf{0.881 $\pm$ 0.007} &  \textbf{0.671 $\pm$ 0.022} &  \textbf{0.917 $\pm$ 0.005} \\
\hline
\end{tabular}
\end{center}
\caption{Classification results of baseline and augmentation models with different settings. Each model is run 5 times for the calculation of all evaluation metrics. 
For fair comparison between ~\cite{xue2019synthetic} and our work, we reimplemented~\cite{xue2019synthetic} for it to use the same pool of synthetic images generated by HistoGAN.}
\label{tb:cervical}
\end{table*}

\begin{figure*}[t]
\begin{center}
  \includegraphics[width=0.99\linewidth]{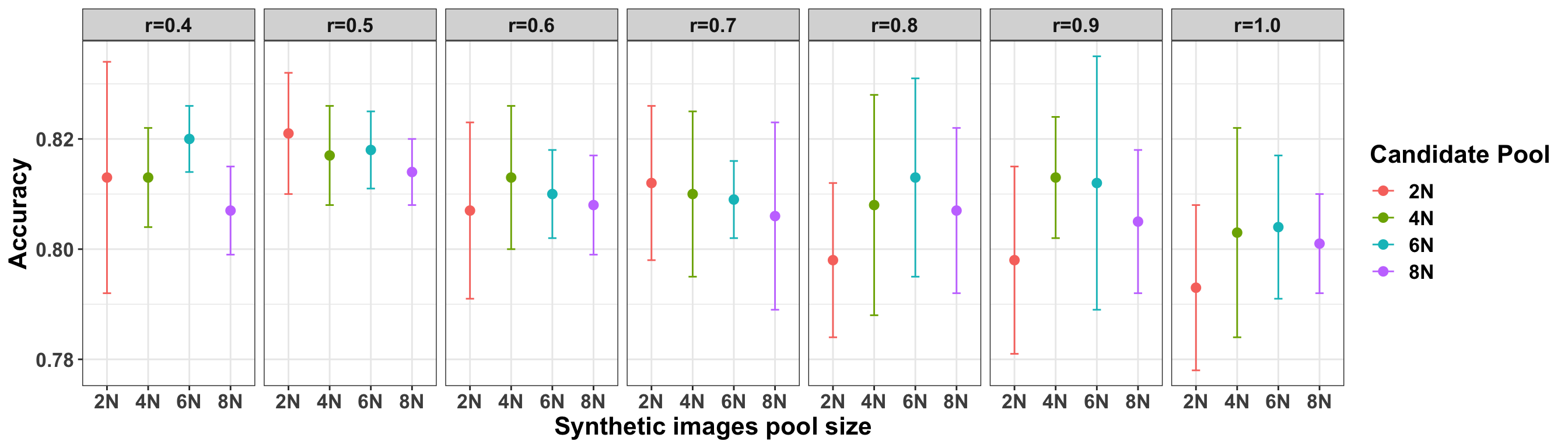}
\end{center}
  \caption{Classification results of the proposed selective synthetic augmentation with different augmentation ratios on the cervical dataset. $N$ in candidate pool sizes indicates the number of images in the original training dataset. For the same candidate pool size, selected images with different ratios are from the same pool. The error bar represents the standard deviation of classification accuracy from 5 multiple runs of each setting, the middle dot refers to the mean of 5 accuracy scores of the aforementioned multiple runs.}
\label{fig:ratio_comparison}
\end{figure*}

\subsubsection{Quantitative Comparisons}
We report quantitative evaluation scores between all baseline augmentation models and our models including the accuracy, area under the ROC curve (AUC), sensitivity and specificity to provide a comprehensive comparison. All models are run for $5$ rounds with random initialization for fair comparison. The mean and standard deviation results of the $5$ runs are reported. 

In Table~\ref{tb:cervical}, we compare quantitative results with different baseline augmentation methods.
We use the same backbone ResNet34 classifier with same hyperparameters setting in all experiments to ensure differences only come from the augmentation mechanisms. Beyond the backbone baseline model ~\cite{he2016deep} without augmentation, we construct a baseline model with traditional augmentation including horizontal flipping and color jittering. Another baseline is GAN augmentation without selection where the training set is expanded by blindly adding GAN-generated images. We also compare the selective augmentation method proposed in this work with our prior work~\cite{xue2019synthetic}.
Since in this work we use HistoGAN, an improved cGAN model that generates better synthetic images (as shown in Fig.~\ref{fig:cervical}) than the cGAN model originally described in~\cite{xue2019synthetic}, we re-implemented~\cite{xue2019synthetic} to also use HistoGAN generated images, for fair comparison of the image selection algorithms. 
From Table~\ref{tb:cervical}, one can see that the selective augmentation algorithm brings obvious benefits to all evaluation metrics, and our full model with augmentation ratio $r=0.5$ achieves best performance in all metrics. More specifically, 
under $r=0.5$, our image selection method improves the classification result by nearly 2\% compared to the method in our prior work~\cite{xue2019synthetic}.   This quantitative result demonstrates that our proposed selection method can better select high-quality images for augmentation than previous work.

To provide further insights on how the choice of the augmentation ratio $r$ affects augmentation performance, we also conduct an ablation study using different values of $r$, on different-sized candidate pools of HistoGAN-generated images. A summary of the ablation study is illustrated in Fig.~\ref{fig:ratio_comparison}. For this study, we generated synthetic image pools of four sizes: $2N$, $4N$, $6N$, and $8N$, where $N$ is the size of the original training set. On these pools, we tested different values of $r$, between $0.4$ and $1.0$. Each test is run for $5$ rounds, and the mean and standard deviation of the $5$ runs are reported. From the results shown in Fig.~\ref{fig:ratio_comparison}, one can see that either too small or too large a value of $r$ compromises the advantage of synthetic augmentation, and the best and most consistent performance gain is achieved at $r=0.5$. This observation is true for all four pools of different sizes. Our explanation for this phenomenon is related to the motivation behind using selective synthetic augmentation: the synthetic images have different levels of quality, and the number of images with good quality and meaningful diverse features generated by a trained GAN model is limited. While our sample selection can provide quality assurance, the total number of diverse, good images that provide complementary information to the existing training set is constrained by the GAN model and more relevantly, by the original labeled training data used to train the GAN model.  Therefore, a larger pool of generated images does not always translate to more high-quality images that will be selected by our method, as shown by this ablation study. Once our selection method has chosen those good images generated by the particular GAN model, adding more images such as images that do not improve diversity but may contain artifacts or bad features would indeed add noise to the training set thus degrade performance.
Since our experiments show that the best augmentation performance is achieved at $r=0.5$, we use this value for all ours and other baseline models and all experiments on the PCam dataset.

In Table~\ref{tb:pcam_all}, we use 3\%, 5\%, 10\% and 20\% of the training data in PCam to simulate training sets with limited annotations and evaluate our models on the full testing set. Compared with the cervical dataset, the baseline classification model achieves higher accuracy on the reduced PCam dataset which makes it more difficult to further improve the performance.  
However, our model still outperforms all baseline models using training sets of different sizes. For instance, when using 10\% of the entire dataset as training data, the classification accuracy improved by $1\%$ when using HistoGAN generated images for augmentation, without selection. After applying image selection, the accuracy is further improved by another $1.7\%$.
By conducting experiments on two histopathology image datasets and showing improved classification performances, we prove that our proposed HistoGAN model and synthetic augmentation algorithm are general and can be applied to various types of histopathology data.

\begin{table*}[t]
\begin{center}
\begin{tabular}{|l|l|c|c|c|c|}
\hline
{PCam} &   Model &      Accuracy &              AUC &      Sensitivity &      Specificity \\

\hline
\multirow{2}{*}{ 3 \%}& Baseline &  0.872 $\pm$ 0.0030 &
0.914 $\pm$ 0.0019 &
0.826 $\pm$ 0.0080 &
0.903 $\pm$ 0.0030 
\\
&\enspace + Selective Augmentation, r=0.5 &  0.900 $\pm$ 0.0024 &
0.933 $\pm$ 0.0016 &
0.865 $\pm$ 0.0060 &
0.924 $\pm$ 0.0030 
\\

\hline
\multirow{2}{*}{ 5 \%}& Baseline &  0.893 $\pm$ 0.0006 &
0.929 $\pm$ 0.0004 &
0.863 $\pm$ 0.0010 &
0.913 $\pm$ 0.0020
\\
& \enspace + Selective Augmentation, r=0.5 &  0.917 $\pm$ 0.0033 &
0.945 $\pm$ 0.0022 &
0.892 $\pm$ 0.0040 &
0.935 $\pm$ 0.0040
 \\

\hline
\multirow{2}{*}{10 \%}& Baseline & 0.910 $\pm$ 0.0012 &
0.940 $\pm$ 0.0009 &
0.883 $\pm$ 0.0050 &
0.929 $\pm$ 0.0030 
\\
&  \enspace + Traditional Augmentation & 0.916 $\pm$ 0.0102 &
0.944 $\pm$ 0.0067 &
0.893 $\pm$ 0.0140  &
0.933 $\pm$ 0.0090 
\\
& \enspace + GAN Augmentation, r=0.5 & 0.920 $\pm$ 0.0020 &
0.947 $\pm$ 0.0014 &
0.898 $\pm$ 0.0050 &
0.935 $\pm$ 0.0020
\\
& \enspace + Selective Augmentation, r=0.5  & 0.937 $\pm$ 0.0011 &
0.958 $\pm$ 0.0007 &
0.916 $\pm$ 0.0070 &
0.951 $\pm$ 0.0040
\\

\hline
\multirow{2}{*}{20 \%}& Baseline &   0.932 $\pm$ 0.0014 &
0.955 $\pm$ 0.0010 &
0.909 $\pm$ 0.0080 &
0.948 $\pm$ 0.0040 
\\
&+ Selective Augmentation, r=0.5  &   0.948 $\pm$ 0.0003 &
0.965 $\pm$ 0.0005 &
0.931 $\pm$ 0.0040 &
0.960 $\pm$ 0.0020 
 \\
\hline
\end{tabular}
\end{center}
\caption{
The performance of baseline and augmentation models using 3\%, 5\%, 10\% and 20\% of PCam as the training set. Each model is run 5 times for the calculation of all evaluation metrics. To further prove the effectiveness of our proposed selective augmentation, we compared the performance of our method and other augmentation methods when using 10\% of PCam as the training set. 10\% is chosen for demonstration because in this case the synthetic images show appealing visual quality as we can observe from Figure \ref{fig:pcam_full}, and consistently the classification performance presents improvement by a large margin.}
\label{tb:pcam_all}
\end{table*}

\section{Discussion}
Our proposed selective synthetic augmentation expands the training dataset by selectively adding synthetic images that do not distort the original data distribution, thus providing quality assurance in augmentation. The selected synthetic images are shown to improve the performance of automated image recognition systems with limited amount of manual annotation. 
We believe our proposed method is applicable to other histopathology image recognition tasks with insufficient annotated data.
In addition, our proposed image selection algorithm is complementary to existing data augmentation methods, which further indicates the generality of our method.

While our selective synthetic augmentation significantly outperforms all baseline models, partial credits should go to the high-fidelity images generated by our proposed HistoGAN. However, the generated images are still not perfect, especially when viewed by expert pathologists, and we expect to further improve our GAN model with help from clinical experts. 
Besides the visual quality of images, the diversity of images also plays a critical role in synthetic augmentation. Since synthetic augmentation is imperative in scenarios with very scarce training samples, combining our pipeline with a GAN model that can learn from limited data~\cite{wang2018transferring,luvcic2019high, noguchi2019image} would further improve the generality of our method. As we provide a solution to assure the synthetic image quality during augmentation, there is still room for improvement in selection mechanisms. More advanced methods for model selection and image selection, such as an end-to-end method and reinforcement learning based method, will be investigated in our future works.

\section{Conclusion}
In this paper, we design a new cGAN model termed HistoGAN for high-fidelity histopathology image synthesis and propose a synthetic augmentation method with quality assurance. By selectively adding realistic samples generated by HistoGAN into the original dataset, our method remarkably boosts the classification performance of baseline models. Experiments on two histopathology image datasets demonstrate the effectiveness and generality of our method.

\section*{Acknowledgments}
This research is supported in part by the Intramural Research
Program of the National Institutes of Health (NIH), National Library
of Medicine, and Lister Hill National Center for Biomedical
Communications.
We gratefully acknowledge the help with expert annotations from Dr. Rosemary Zuna of the University of Oklahoma Health Sciences Center. We also thank Dr. Joe Stanley of Missouri University of Science and Technology for making the cervical histopathology data collection available.

\bibliographystyle{model2-names.bst}\biboptions{numbers}
\bibliography{refs}

\end{document}